\documentclass[conference]{IEEEtran}
\IEEEoverridecommandlockouts
\usepackage{cite}
\usepackage{amsmath,amssymb,amsfonts}
\usepackage{algorithmic}
\usepackage{graphicx}
\usepackage{textcomp}
\usepackage{xcolor}

\def\BibTeX{{\rm B\kern-.05em{\sc i\kern-.025em b}\kern-.08em
    T\kern-.1667em\lower.7ex\hbox{E}\kern-.125emX}}
\usepackage{fancyhdr}
\begin{document}
\title{Requirements Analysis of Variability Constraints in a Configurable Flight Software System
\thanks{This material is based upon work supported by the National Science Foundation under grant CCF\#2211589.}}

\author{\IEEEauthorblockN{Chin Khor}
\IEEEauthorblockA{\textit{Computer Science Department} \\
\textit{Iowa State University}\\
Ames, USA \\
chinkhor@iastate.edu}
\and
\IEEEauthorblockN{Robyn R. Lutz}
\IEEEauthorblockA{\textit{Computer Science Department} \\
\textit{Iowa State University}\\
Ames, USA \\
rlutz@iastate.edu}
}

\maketitle
\thispagestyle{fancy}
\begin{abstract}
Variability constraints are an integral part of the requirements for a configurable system.  The constraints specified in the requirements on the legal combinations of options define the space of potential valid configurations for the system-to-be. This paper reports on our experience with the variability-related requirements constraints of a flight software framework used by multiple space missions.  A challenge that we saw for practitioners using the current framework, now open-sourced, is that the specifications of its variability-related requirements and constraints are dispersed across several documents, rather than being centralized in the software requirements specification.  Such dispersion can contribute to misunderstandings of the side-effects of design choices, increased effort for developers, and bugs during operations.   Based on our experience, we propose a new software variability model, similar to a product-line feature model, in the flight software framework.  We describe the structured technique by which our model is developed, demonstrate its use, and evaluate it on a key service module of the flight software.   Results show that our lightweight modeling technique helped find missing and inconsistent variability-related requirements and constraints.  More generally, we suggest that a variability modeling technique such as this can be an efficient way for developers to centralize the specification and improve the analysis of dispersed variability-related requirements and constraints in other configurable systems.  

\end{abstract}

\begin{IEEEkeywords}
Requirement analysis, Variability constraints, Variability requirements, Configurable system, Feature model
\end{IEEEkeywords}

\section{Introduction}

Variability constraints are an integral part of the requirements for a configurable system \cite{batory2021, oPoBoLi05}.  The constraints specified in the requirements on the combinations of options that are allowed/disallowed define the space of potential valid configurations for the system-to-be \cite{NadiBKC15}. This paper reports our experience developing a variability model to specify and analyze the variability-related requirements and constraints of a NASA configurable flight software framework used by multiple space missions.  

A challenge that we saw for practitioners in using the current flight software framework is that specification of its variability-related requirements and  constraints are dispersed across several documents rather than being centralized in the software requirements specification.  This dispersion can contribute to misunderstandings of the side-effects of design choices, increased effort for developers, and bugs during operations.

Such dispersion is a persistent problem especially for configurable systems and product lines, in that information about variability-related requirements, such as options, dependencies and constraints on valid configurations, is often scattered and thus easily overlooked \cite{CashmanCRC18}.  While much progress has been made in formal modeling and analysis of all valid configurations \cite{apel2014feature, SoaresSMA18}, this research tends not to meet projects ``where they are," in that most projects do not have or want formal models.  Consequently many projects do not find variability constraints that are missing from the requirements, are inconsistent, or are in conflict until testing or operations.  

Based on our experience with the configurable flight software requirements, we propose addition of a new software artifact, a variability model, to the existing flight software framework. A variability model is a higher-level representations of options and constraints \cite{BergerN15}.  Our construction is based on a product-line feature model and is intended to be useful to future mission developers of the flight software framework. It integrates the scattered variability requirements and constraints into a coherent, centralized, variability-aware model for requirements analysis. 

In this paper we describe our new, lightweight, tool-supported variability modeling technique, VarCORE, and discuss our experience applying it to a key service module of the flight software.  Results reported here show that the variability model helped find missing variability-related requirements and constraints, inconsistencies among the specified variability requirements and constraints,  and variability requirements and constraints not implemented in the code. 

We anticipate that this new  modeling technique, supported by existing open-source tools, will be useful to future users of the flight software framework in both the requirements engineering and aerospace communities. More generally, we suggest that a variability model such as this can be an efficient way to centralize the specification of variability-related requirements constraints that are dispersed among documents for other configurable systems. 

The contributions of the paper are three-fold:
\begin{itemize}
\item A proposed new variability model for the NASA configurable flight software framework that integrates, specifies and analyzes variability-related requirements and constraints that are currently dispersed among documents 
\item Analysis results identifying missing and inconsistent variability requirements and constraints for repair
\item A structured and partially automated technique, VarCORE, to develop and analyze the variability model for potential use in other configurable systems.  
\end{itemize}

A fourth, indirect contribution of the paper is that it describes an open-sourced, configurable flight software framework that we believe merits increased use as a platform for requirements engineering innovations.   To the best of our knowledge, it has not been used by the requirements engineering community, although it has been used in both architecture and testing papers \cite{ganesan2009verifying, ganesan2013analysis, miranda19}. Greater use of available, real-world systems such as this flight software framework benefits both requirements researchers and practitioners.  Evaluation of new RE techniques on such systems can provide both real-world feedback needed to improve them and the evidence needed for their ready uptake by industry.  

The rest of the paper is organized as follows.  Section 2 describes the context of the NASA configurable flight software framework and the project's variability requirements related practices.  Section 3 presents the problem statement and challenges.  Section 4 describes our approach in VarCORE to creating a variability model for the flight software.  Section 5 presents and discusses our experience applying VarCORE to a part of the flight software and its analysis findings. Section 6 discusses lessons learned and their generalizability, and related work. Sect 7 provides concluding remarks. 

\section{Industrial Context}
\label{context}

The core Flight System (cFS)  \cite{cFSurl}, \cite{mccomas2016core} is a configuration-managed, software framework of flight software for spacecraft.  It contains a Platform Support Package (PSP), Operating System Abstraction Layer (OSAL), Core Flight Executive (cFE) Services, cFS Applications and Tools, and supporting artifacts. The cFS applies a software product line approach to reuse across missions.  Commonalities and variabilities from a rich heritage of space missions at the NASA Goddard  Space Flight Center are implemented in the form of mission-independent flight software \cite{nasa21}.  
The cFS is feature-based, meaning that various combinations of features (units of functionality or configuration options) are used to meet different needs \cite{Lamsweerde09, BergerLRGS0CC15, ApelBatory13, RheinGAS0B15}. The objective is to provide high-quality, reusable core flight software that can meet the majority of basic flight requirements for a variety of spacecraft missions. 

The core Flight System was open-sourced in 2015 to further the business goal of reducing time and cost for developing new flight software in spacecraft projects adopting the cFS.  The second author served on the software architectural review board for the core Flight System, tasked with providing a product-line perspective, in preparation for its being open-sourced. In 2020, cFS was awarded NASA’s Software of the Year. At NASA it has been used on at least nine spacecraft projects and it is also chosen by NASA Goddard Space Flight Center (GSFC), Johnson Space Center (JSC) and Johns Hopkins University Applied Physics Lab (APL) for all future embedded flight software projects \cite{cFSurl}. As part of the Lunar Gateway program, work is also underway to certify cFS as suitable for human-rated vehicles \cite{ngo21}.  According to the team lead for certification, “We work on maybe two or three missions a year, but outside of NASA, people are trying it out, finding new ways to use it, and making suggestions for improvement” \cite{nasa21}. Examples include the CubeSat nanosatellites and small spacecraft \cite{mccomas2016core}. Educationally, OpenSatKit (OSK) adopts cFS to provide a free flight software system platform for aerospace and STEM education \cite{mccomas2021opensatkit}. 
\begin{figure} [tb]
\centerline{\includegraphics[scale=0.65]{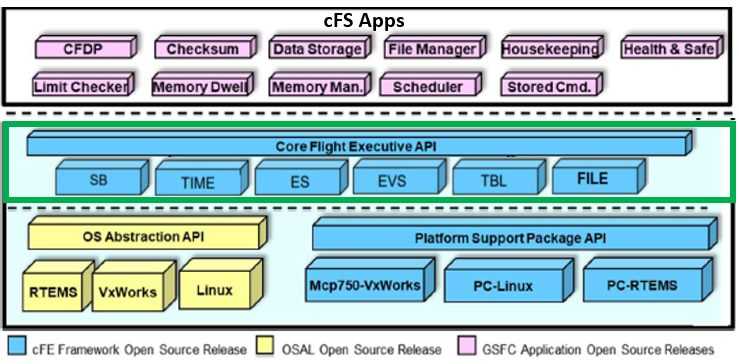}}
\caption{cFS Software Architecture and Organization. Excerpt from \textit{Core Flight System (cFS) Training}, p. 9  \cite{timmons2020core}.}
\label{cFSArch}
\vspace*{-3mm}
\end{figure}
\vspace*{-1mm}
\subsection{Architectural Design for Variability Support}
The cFS was designed with reusability and variability in mind. Figure \ref{cFSArch} shows the layered modular architecture of the cFS. At the bottom layer, the cFS supports multiple hardware platforms (including i686/x86 based PC, ARM-based PC, powerPC-based PC and MCP750) and operating systems (including Linux, VxWorks and RTEMS) via abstractions by OSAL and PSP. It thus hides the implementation details of hardware and operating systems from the applications and executive.  

The core Flight Executive (cFE) provides core services. At the top layer, the cFS defines standardized core Application Programmer Interfaces (APIs), which are the only interfaces for applications to access the cFE core services.   
Although the cFE core services are common across most flight system projects, the cFE does provide configurable parameters to determine variations of the core services at compilation. To change core services' configuration, the entire cFE must be recompiled and restarted with a system reset. 

The cFS is designed to support plug-and-play applications using these standardized APIs.  This means that applications can be loaded and removed at runtime. Applications define mission functionality with the support of the cFE. The cFS architecture ensures that different combinations of applications can be run independently and concurrently to fulfill mission requirements.

\subsection{Strategies for Variability Management}
There are three strategies for managing software product variability in the cFS. First, the cFS uses its architecture design to manage mission-specific application variability. Each application is an independent entity. To establish connection with the cFE, each application is required to register and subscribe for services' access to cFE. Inter-application interaction is only supported via cFE, i.e. direct interaction is constrained. Ganesan et al. \cite{ganesan2009verifying} analyzed the cFS product line architecture and verified that there is no inter-dependency and direct interaction among applications, complying with architectural design for the cFS application layer. 

Second, the cFS uses its build system to manage hardware variants, OS variants, and module and test  configurability. The cFS build system checks for configurable parameters 
during compilation and builds the flight software product accordingly. Thus, applications and cFE source code can be run on different hardware and OS configurations via the support of the abstraction layer and defined APIs.  

Third, the cFS uses conditional compilation C pre-processor (CPP) directives (such as \textit{\#if, \#ifdef, \#elif, \#ifndef}) to compile the code based on the setting defined by configurable parameters. There are 139 configurable parameters \cite{mccomas2016core} defined in the cFE 6.7.0 release that are handled by conditional compilation. We found that only 15 of the configurable parameters represent variant features for configuring the system functional behavior. The rest of the configurable parameters are mainly defined to represent  system (i.e., platform and mission) properties such as memory size, message id, event type, etc. The cFE uses CPP directives to verify the validity of these system properties before system configuration.  

\section{Problem Statement: Variability-Aware Requirements Constraints}
\begin{figure} [b]
\centerline{\includegraphics[scale=0.7]{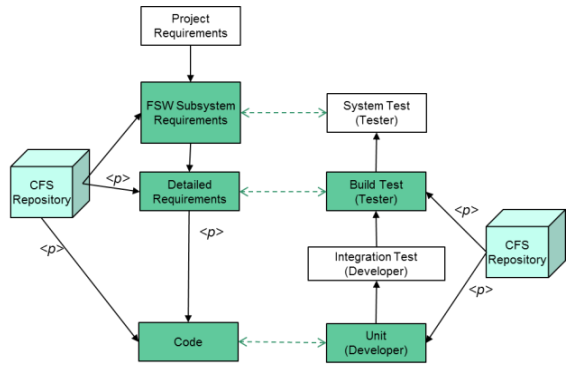}}
\caption{V-Model of cFS-based Software Development Lifecycle. Reprinted from \cite{mccomas2016core}. The shaded components are cFS artifacts and $<$\textit{p}$>$ notation indicates a parameterized artifact. }
\label{Vmodel}
\end{figure}
The cFS describes their requirements strategy \cite{CFEReq} as incorporating requirements engineering best practices  to manage software variability. The requirements are parameterized to support various missions. Fig. \ref{Vmodel} shows the software engineering V-Model for the cFS-based project life cycle. On the left, the requirements are refined from high-level system requirements into lower-level detailed functional requirements for flight software development. On the right, corresponding tests are created to verify/validate the requirements at each level. In addition, the functional requirements are written using recommended practices to assure quality \cite{ieee1998ieee}. Any change in requirements requires review and approval before becoming effective. 

Nevertheless, there remains a risk that variability constraints  may be overlooked during requirements refinement, especially if selections of options are delayed until design or compilation phases. The cFS uses pre-processor directives to verify and assure the consistency of variability requirements and constraints at build time. However, we observed that the constraint verification is done independently without explicit traceability to the variability requirements. Moreover, it is known that the gap between requirements specification and implementation tends to grow \cite{lethbridge2003software} if requirements updates are not performed in a timely manner.   

Another challenge that we observed with the current framework is that the variability-related requirements are not specified in a single requirements specification document. Instead they are dispersed across several documents. Some constraints are mentioned but not described, or are described but not specified as requirements. Such dispersion can complicate the requirements analysis and thus increase the effort for variability design, implementation and testing.

The cFE Flight Executive Software Requirements Specification is relatively short (30 pages plus appendices).  For example, it contains only 32 requirements for the six cFE services.  The other requirements document, the Functional Requirements table, has 398 brief (one-line) ``shall" textual requirements.  It does not document the configuration options available nor the intra-cFE variability constraints that the developers of a new product need to know. On the other hand, the cFE user guide is very long (1,056 pages). It specifies requirements for the configuration options and the variability constraints. However, its length renders it difficult to navigate and use. 

The dispersion of variability-related requirements and constraints among documents creates a Goldilocks problem \cite{goldilocks}, named after the children's story in which Goldilocks finds the Papa Bear's  chair too big and the Mama Bear's chair too small, but the Baby Bear's chair is ``just right" for her.  The analogy refers to the challenge that developers of a new system using the cFE face in currently not having a central ``just right" sized specification. 

Towards making it easier for developers to understand the variability features and constraints on the feature combinations, {\em we built a new, visual variability model} that captures this requirements information in a standardized and familiar graphical form.  We do not claim that it is Goldilocks-optimal; however, show that it facilitates discovery of missing and inconsistent variability-related requirements and constraints.      

\section{Approach}\label{Approach}

\begin{figure}
\centerline{\includegraphics[scale=0.65]{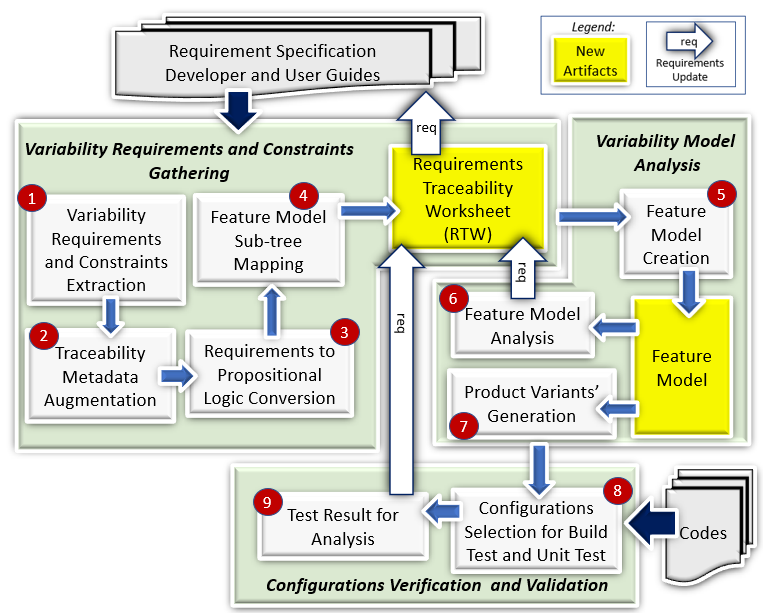}}
\caption{VarCORE Overview.  The new artifacts produced by the variability-related requirements and constraint analyses are highlighted in yellow. 
}
\label{ProposedFrameworks}
\end{figure}

In this section we describe our VarCORE technique and the new software
variability model it produces, proposed for inclusion in the flight software framework. Fig. \ref{ProposedFrameworks} gives an overview of VarCORE and the artifacts it uses and provides for variability-related requirements and constraints analysis in a configurable flight software. The circled numbers in the figure map to the elements in the description below. 

There are three components in the VarCORE process: (1) Variability Requirements and Constraints Gathering, (2) Variability Model Analysis, and (3) Configurations Verification \& Validation. 

\subsection{Variability-Related Requirements and Constraints Gathering}\label{VGathering}
The primary input to VarCORE is the software requirement documents. For large-scale software projects there are typically more than one requirements document, starting from system requirements and decomposing them into subsystems, modules and units requirements. Often these are produced by different engineering teams at different levels of refinement.  This can create a challenge to maintaining consistency and traceability among all requirement documents.

\begin{table}[b]
\caption{Requirements Traceability Worksheet (RTW).}
\centerline{\includegraphics[scale=0.45]{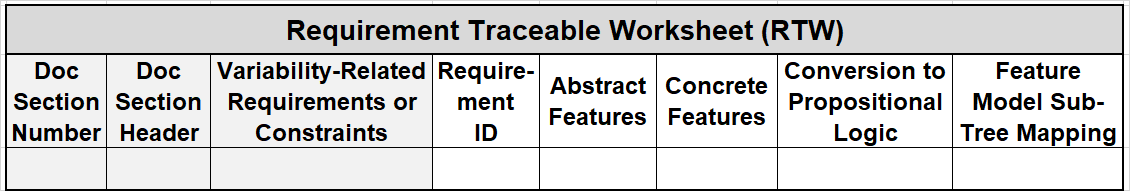}}
\label{reqWorksheet}
\end{table}

In our approach, we propose a lightweight process to gather the scattered variability requirements and constraints from diverse requirement documents into a centralized 
Requirements Traceability Worksheet (RTW), illustrated in Table \ref{reqWorksheet}.  Capturing in the RTW the dispersed information needed to construct the feature model enables the step-by-step construction of a feature model that is consistent with the documented variability-related requirements and constraints.

First, the variability requirements and constraints are extracted from the requirement documents and entered into the RTW, one requirement/constraint per entry. Then, we identify the relevant terms such as the variant features and/or the feature constraints from each requirement/constraint, and name them as concrete features which are key elements of a feature model. In implementation, these terms are generally the configurable parameters that manage variability or establish constraints in the code. As an example, ``\textit{cFE is configured to function as a time server or time client}" is an identified variability requirement. Clearly  \textit{time server} and \textit{time client} are two configuration variants of the cFE function. Accordingly, we define \textit{time server} and \textit{time client} as concrete features, and group them under the \textit{cFE function}. Usually concrete features are terminal nodes of a feature model. However, we chose to allow concrete features in non-terminal nodes to keep the feature tree smaller. Our feature tree still can be transformed into a GenVoca feature tree\cite{batory2021}; that transformation adds nodes, so we avoid it here. 

Second, we assign a unique requirement ID as an identifier for each entry in the RTW. Then we augment the traceable metadata when naming the abstract feature \cite{batory2021,thum2011abstract, Cleland-HuangHHLM12}. In the context of a feature model, the abstract features are non-terminal nodes used to structure the feature diagram by grouping the related concrete features (identified in the previous step), so the name of an abstract feature should be representative for grouping concrete features. Our naming convention is to augment the abstract feature name (e.g. ``name"\_``id") with traceable metadata that is extracted along with the requirement/constraint. The type of metadata is varied depending on the requirement structure. If the requirements/constraints are specified in the structured 
requirement document, the metadata can be the requirement ID and the heading description for each 
requirement/constraint, e.g., Time\_Function\_T1001. On the other hand, if the requirements/constraints are extracted from a free-form textual requirement document, the metadata can be document section number and 
heading, e.g., Time\_Server\_Client\_1\_22\_4\_5. Although abstract features are essential elements for a feature model, they do not contribute to product variant generation as they do not appear in the code. 

Third, after identifying the abstract and concrete features
for each variability requirement/constraint, VarCORE translates the textual variability requirements and constraints into propositional logic expressions\cite{epstein2013semantic} using the propositional logic forms in Table \ref{ModelRule}. For example, ``\textit{cFE is configured to function as a time server or time client}" will be translated to \\
\hspace*{2cm} Time\_Function\_T1001 $\iff$ \\
\hspace*{0.1cm} $(\textit{time server} \wedge  \neg \textit{time client}) \vee (\neg \textit{time server} \wedge  \textit{time client})$ \\
This translation is done manually, and we discuss the limitations of NLP techniques that drove our decision not to automate this in related work in Section \ref{relatedwork}. 
The translation to propositional logic form serves two purposes. One is to help verify the correctness of the variability requirement or constraint. A requirement or constraint statement in natural language can be ambiguous and interpreted differently by different developers. Translating it to propositional logic which has only one mathematical interpretation helps verify the correctness of the requirement or constraint. 

Fourth, once in propositional logic form, the relationship between abstract and concrete features can be determined using the defined mapping rules \cite{batory2021} based on Feature Oriented Domain Analysis (FODA) notation \cite{kang1990feature} (see Table \ref{ModelRule}). Features are units of functionality, i.e., functional requirements \cite{Lamsweerde09, BergerLRGS0CC15, ApelBatory13}. These rules convert abstract features, concrete features and their relationships to feature model sub-trees. For instance, the propositional logic expression  ``\textit{cFE is configured to function as a time server or time client}" is mapped to rule R4, which states that a \text{Time\_Function\_T1001} (abstract feature) can select only one child (concrete feature), either \textit{time server} or \textit{time client}. By following the rules given in Table \ref{ModelRule}, we can systematically convert all variability requirements and constraints in the RTW to their associated feature model sub-trees. Table \ref{processedReq} shows a sample of the RTW generated for the Time service module of the cFE.

\begin{table}[tb]
\caption{Rule Definition for Mapping Propositional Formulas to Feature Model Sub-trees.}
\centerline{\includegraphics[scale=0.5]{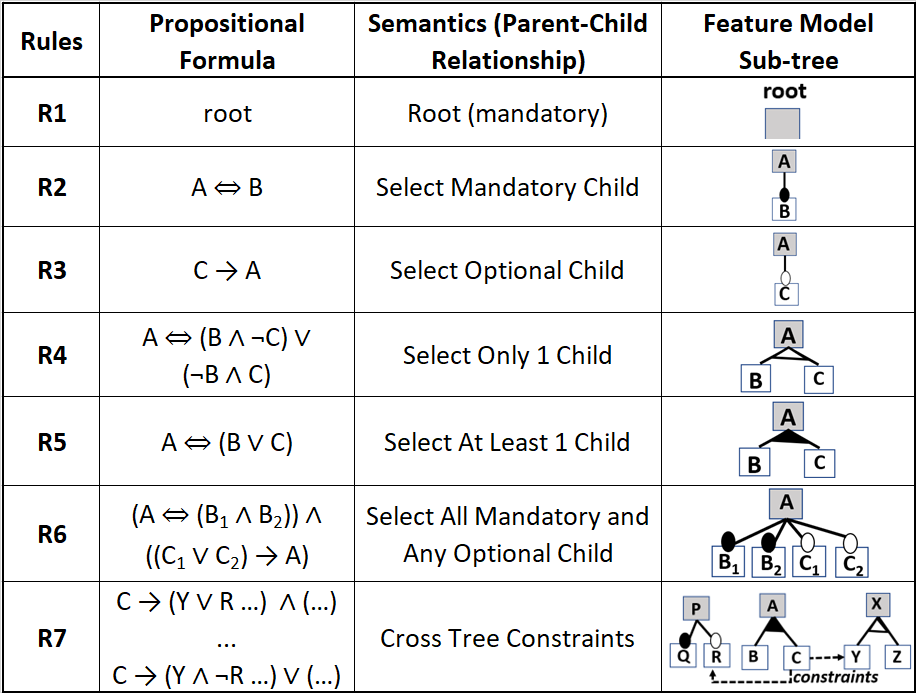}}
\label{ModelRule}
\end{table}

The RTW is composed of  entries for variability requirements and variability constraints. Two variability requirements are in a hierarchical relationship if the feature of one variability requirement is the parent node of the feature (i.e., child node) of another variability requirement. Thus, a hierarchical relationship between two variability requirements is also a parent-child relationship. A variability constraint is represented by another type of relationship, the cross-tree constraint, which is an implication relationship (i.e., if-else constraint) between two or more concrete features across variability requirements.  

Evaluation to assure consistency among variability requirements and constraints is not a trivial task. VarCORE creates a graphical variability model (using feature modeling) to aid the variability requirements and constraint analysis (see next section). Additionally, to handle requirement changes, entries can readily be added, modified, or deleted in the RTW at any time to create updated or experimental variability models for comparison, analysis, or troubleshooting.

\begin{figure*}
\centerline{\includegraphics[scale=0.85]{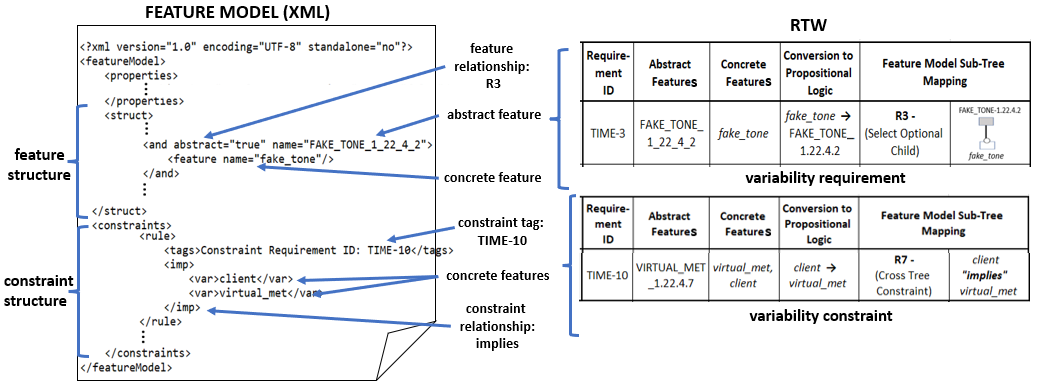}}
\caption{Concatenating Feature Model Sub-trees using XML Template for Feature Model Construction.}
\label{rtw2xml}
\end{figure*}

\subsection{Variability Model Analysis} \label{VAnalysis}
Once the first new artifact, the RTW, is constructed, we use it to generate a second new artifact,  a feature model that represents the variability requirements and constraints in a graphical view. A feature model is a standard type of variability model, with table-based models being another common type. A feature model is a feature diagram, described below, together with the constraints on those features \cite{batory2021}, \cite{kang1990feature}. 

In VarCORE we employ the open-source tool FeatureIDE, which has been widely used for feature modeling \cite{meinicke2017mastering} to construct the feature model (refer to the fifth step in Fig. \ref{ProposedFrameworks}). We formulate each feature model sub-tree (from RTW) in the Extensible Markup Language (XML) format recognized by FeatureIDE. Within a sub-tree, the hierarchical relationship between abstract and concrete features must be maintained. The XML representations of the sub-trees then are concatenated and integrated into an XML file for the feature model. Fig. \ref{rtw2xml} shows the process of concatenating variability requirement and constraint sub-trees into the XML template recognized by FeatureIDE. The order of concatenation for sibling sub-trees and constraints does not impact the feature model computation, but does decide features' and constraints' relative display position in the feature diagram. We developed a program\footnote{github.com/chinkhor/VarCORE/tree/main/scripts/RTW2FeatureModel} to automatically construct a feature model (XML format) from the structured RTW entries. Once the feature model in XML format is created, we input it to FeatureIDE to generate a graphical representation of the feature model automatically (see Fig. \ref{TimeFeatureModel} for the final feature model generated for the cFE Time service module).

\begin{figure*}
\centerline{\includegraphics[scale=0.6]{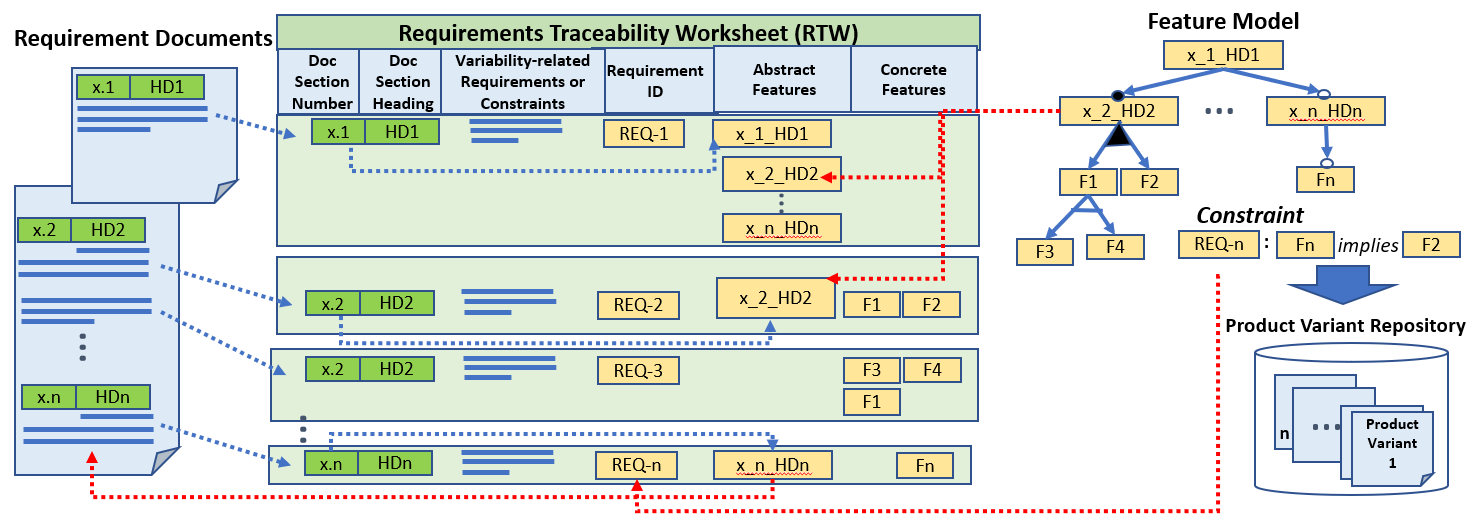}}
\caption{Traceability among Requirement Documents, RTW and Feature Model.}
\label{traceability}
\end{figure*}

We use FeatureIDE's built-in capabilities to analyze the model's validity, including checks for void-model detection (i.e., feature model without any valid configuration), dead-feature detection (i.e., a feature that is not part of any valid configuration), false-optional feature detection (i.e., optional feature but present in all products), constraint redundancy, and 
conflict of constraints. During the feature model analysis, FeatureIDE flags any occurrence of these errors in the feature diagram and reports them in its statistical report. Using our RTW, we then can trace any reported inconsistency or error both back to the corresponding variability requirement or constraint, and from there to its source in a requirement document.  Fig. \ref{traceability} shows the traceability among the requirements documents, the RTW and the feature model.   

After completion of the feature model analysis, FeatureIDE can generate all valid configurations (or product variants) encoded in the feature model. These valid configurations are then exported to a repository for our subsequent use. 

\subsection{Verifying and Validating Configurations} 
\label{VTesting}

The feature model undergoes static analysis to check the consistency among the variability requirements and constraints. However, the analysis does not detect variability bugs in the implementation.  To detect inconsistencies between the specification of the variability requirements and constraints in the feature model and the source code, we use  
the generated product variants from our feature model as inputs to the cFS build system and unit test framework for variability bug detection. 
Each product variant represents a valid configuration that satisfies all variability
requirements and constraints specified in the RTW. Verifying these product variants via build tests and unit tests helps assure consistency between the variability requirements/constraints and the code that should satisfy them, while inconsistencies could indicate flaws in the requirements. 

The product variant files generated by FeatureIDE are in XML format. They thus must be translated to configuration files that are recognized by the build system. To achieve this, we parse and convert the concrete features in XML to the corresponding variable names in the source code. This step uses a pre-defined mapping table between concrete feature names and code variable names. The build system then modifies the configuration setting as specified by the product variant configuration file to start the build test and unit test. If the configuration space is small, all product variants can be selected for testing. Otherwise, random or targeted sampling of product variants for testing can be done. 

Finally, the test results for each product variant is gathered for analysis. Any build failure or unit test error is an indication of potential inconsistency between the implementation and the variability requirements/constraints. Once the variability bug is identified, the RTW can be used to trace the source of variability requirements and constraints which are related to the inconsistency for further analysis. 

\section{Application and Findings} \label{App_Findings}

In this section we describe our application of VarCORE on a portion of the cFS flight software. We also present our findings regarding the following three 
research questions:
\begin{itemize}
\item RQ1: Are there missing variability-related requirements and/or constraints?
\item RQ2: Are there inconsistencies among the specified variability requirements and/or constraints?
\item RQ3: Are there variability requirements and/or constraints not implemented in the code?
\end{itemize}

\begin{table} [tb]
\caption{TIME Configuration Variability and its Functionality.}
\vspace*{-3mm}
\begin{center}
\begin{tabular}{|c|l|}
\hline
\textbf{Variability} & \textbf{Functionality} \\
\hline
\textit{server} &  Time operation in Server mode \\
\hline
\textit{client} & Time operation in Client mode \\
\hline
\textit{big\_endian} & Force tone message in big endian order \\
\hline
\textit{virtual\_met} & Configure as virtual MET if there is no local \\ 
& hardware MET. \textit{virtual\_met = false} indicates \\
& local hardware MET is enabled. \\
\hline
\textit{source} & Source of time data is external\\
\hline
\textit{source\_met} & Type of external time data source is MET \\
\hline
\textit{source\_gps} & Type of external time data source is GPS \\
\hline
\textit{source\_time} & Type of external time data source is spacecraft time \\
\hline
\textit{signal} & Support primary and redundant tone signal selection \\
\hline
\textit{tai} & Default time format is International Atomic Time \\
\hline
\textit{utc} & Default time format is Coordinated Universal Time \\
\hline
\textit{fake\_tone} & Enable fake tone signal generation in the absence \\
& of real hardware signal \\
\hline
\textit{at\_tone\_was} & Tone signal arrives before "time at the tone" data \\
\hline
\textit{at\_tone\_will\_be} & Tone signal arrives after "time at the tone" data \\
\hline
\end{tabular}
\label{TimeVar}
\end{center}
\vspace*{-4mm}
\end{table}

\subsection{Application}
We chose to evaluate our VarCORE technique on the cFE Time service module (TIME) since 14 of the 15 variant feature configurable parameters that define cFE system behavior are used for Time configuration. TIME is the core service that provides spacecraft time correlation, distribution, and synchronization services. These encompass multiple variabilities, including the time operational mode (server or client mode), time format, source of time, scheme for tone signal, etc. Table \ref{TimeVar} shows all 14 TIME configuration variabilities and their corresponding functionalities. The configuration of these variabilities is defined at build-time and dictate the TIME system behavior at run-time. 

Initially we selected the cFE Software Requirement Specification (cFE requirements.docx) and the cFE Funtional Requirements specification (cFE FunctionalRequirements.csv) as the sources from which to gather the variability-related requirements and constraints for TIME. However, we found that only six variability requirements are described in these two requirements documents.  We then discovered that, instead, the cFE User's Guide (which is the reference for application, tool and test development) describes the TIME variability requirements and constraints in detail. We thus used it as one of the primary input documents for our VarCORE application and evaluation.  

\begin{table*} [tb]
\caption{Excerpts from Requirements Traceability Worksheet (RTW) for cFE Time Configuration. Variability requirements and constraints that are unsatisfiable or conflicted with others are highlighted in red.  The full RTW is available at https://github.com/chinkhor/VarCORE.}
\centerline{\includegraphics[scale=0.5]{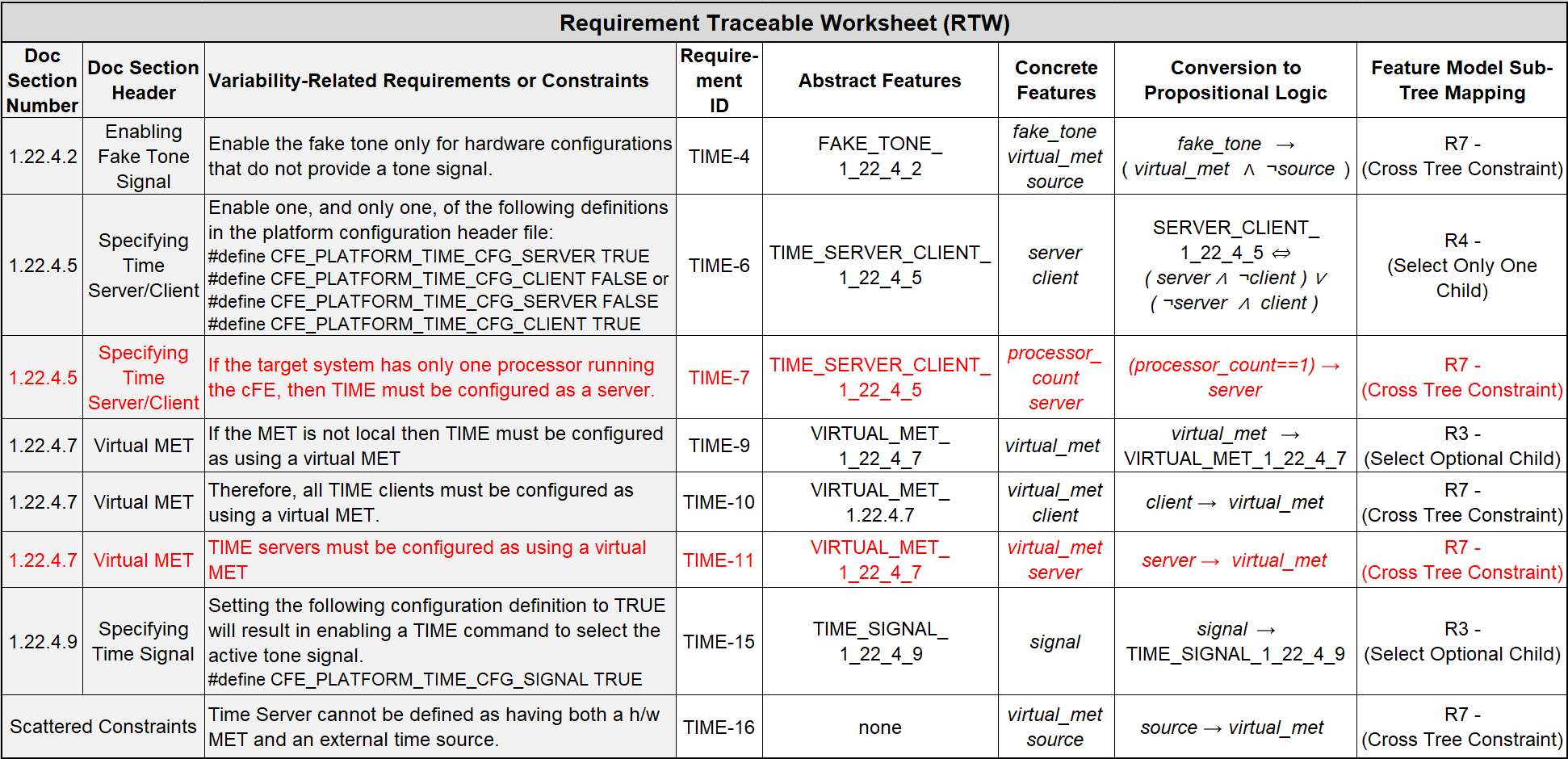}}
\label{processedReq}
\end{table*}

\subsection{Findings} \label{findings}
Table \ref{processedReq} shows an excerpt of the first generated artifact produced with VarCORE, the RTW (Requirements Traceability Worksheet) described in Section \ref{VGathering}.  Following the process described there, 10 variability requirements and 7 variability constraints were identified for TIME configurations and extracted from the cFE User's Guide. Then, 9 abstract features and 15 concrete features were derived from the 17 variability requirements and constraints. 

With TIME's variability\textminus related requirements and constraints now gathered into the structured RTW, we translated each requirement or constraint into its propositional logic formula. In parallel, we reviewed all the formulas to verify the correctness and consistency of the variability requirements and constraints for the TIME configuration. 

Interestingly, in the review process we discovered that one cross-tree constraint could never be satisfied. TIME\textminus7 (see Table \ref{processedReq}) could not be satisfied as there was no definition for processor count, neither in TIME,  nor in PSP or  OSAL (we manually checked PSP and OSAL later). TIME\textminus7 thus was flagged as a unsatisfiable variability constraint and highlighted in red on the RTW. 

After verification of the variability requirements and constraints was completed, we created the feature model for TIME using the FeatureIDE tool (described in Sect. \ref{VAnalysis}) using the satisfied variability requirements and constraints but without TIME\textminus7. The graphical feature model is the second artifact produced by VarCORE, for use in the feature model analysis. 

\begin{figure*}
\centerline{\includegraphics[scale=0.6]{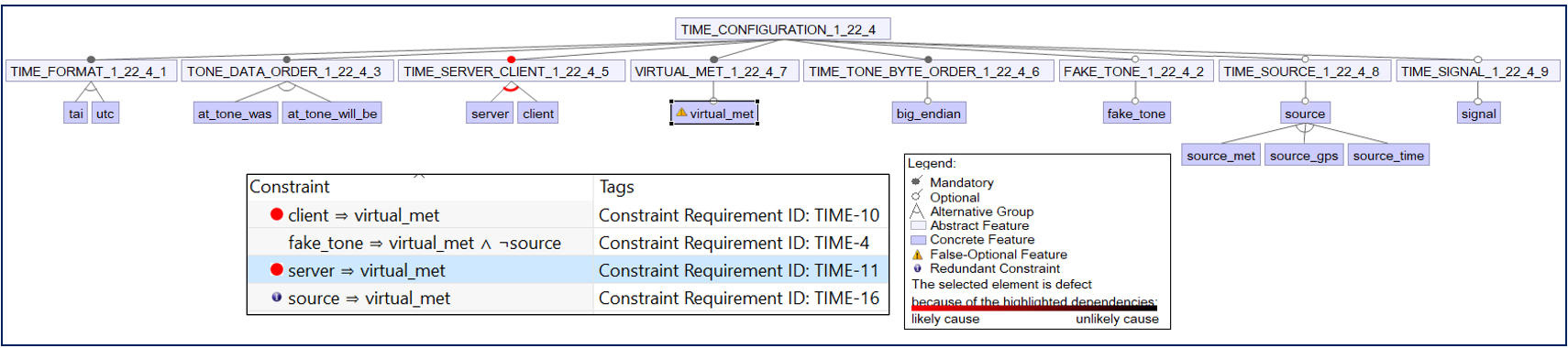}}
\caption{Initial Constructed Feature Model Showing Detection of Inconsistent Constraints.}
\label{InvalidFeatureModel}
\end{figure*}

Fig. \ref{InvalidFeatureModel} shows the {\em initial} constructed feature model, including the cross-tree constraints, for TIME that was generated by the FeatureIDE tool. However, FeatureIDE reported the existence of a false-optional feature in the initial feature model.  It showed that this was caused by the constraints specified by TIME\textminus10, TIME\textminus11 and TIME\textminus16, as shown in the figure. 

\begin{figure*}
\centerline{\includegraphics[scale=0.6]{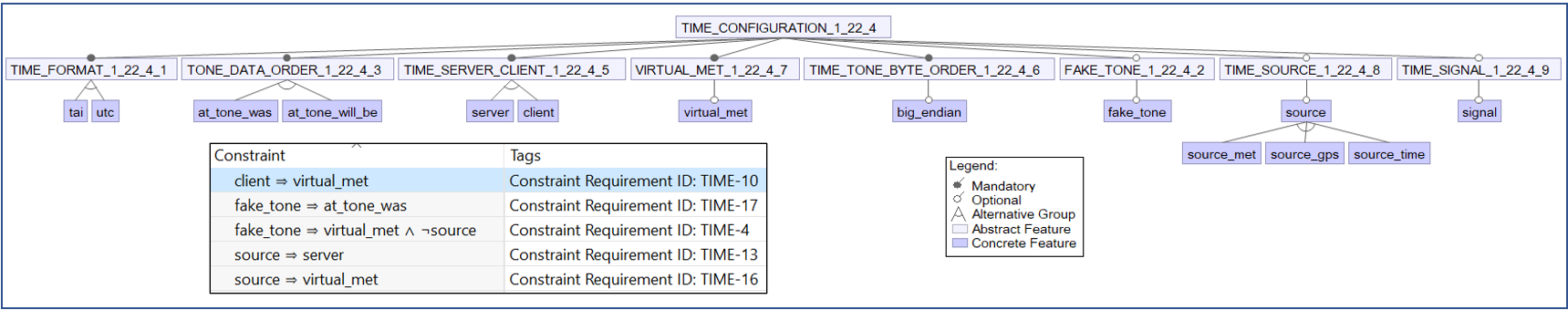}}
\caption{Final Feature Model for Time Service Configuration. The full feature model is at https://github.com/chinkhor/VarCORE.}
\label{TimeFeatureModel}
\end{figure*}

\begin{figure} [bt]
\centerline{\includegraphics[scale=0.8]{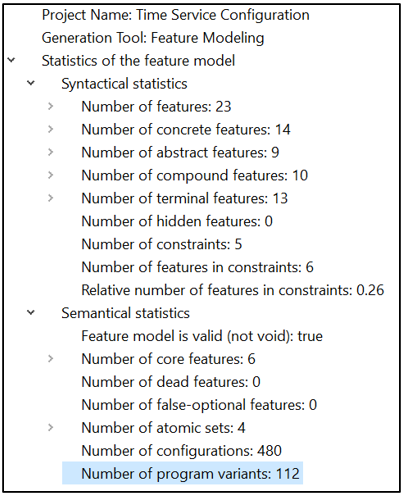}}
\caption{Feature Model Analysis Statistics by FeatureIDE}
\label{FeatureModelStat}
\vspace*{-3mm}
\end{figure}

Further investigation  by the authors indicated that the culprit was TIME\textminus11, as this constraint would prohibit the Time Server from using the local hardware's MET (Mission Elapsed Time). The variability constraint specified in TIME\textminus11 was thus judged to be erroneous and was highlighted in red in the RTW for attention by the developers. Using the traceability provided by the feature model and RTW, we could readily identify the source of this requirement specification, so that it could be repaired. 
Figs. \ref{TimeFeatureModel} and \ref{FeatureModelStat} show the updated result, a valid feature model, after the variability constraint TIME\textminus11 was removed to resolve the inconsistency. 

After completion of the model analysis, we generated all 112 product variants for verification and validation using cFS's build system and unit test framework. Each product variant represents a valid configuration supported by the feature model. Our goal is to assure that the cFE does fulfill the variability requirements and constraints specified in the RTW. We transformed all 112 product variants (in XML files) to configuration files recognized by the cFS build system using the mapping table shown in Table \ref{var2code}. The list of C pre-procesor (CPP) directives for TIME was retrieved using the Linux ifnames tool, and the mapping was straight forward as there were common keywords between the concrete features and the CPP variable names. 

\begin{table} [tb]
\caption{Mapping between Concrete Features and Code.}
\vspace*{-3mm}
\begin{center}
\begin{tabular}{|c|c|}
\hline
\textbf{Concrete Features in}& \textbf{C Pre-processor Directives in} \\
\textbf{Feature Model}& \textbf{CFE Time Service Module (Codes)} \\
\hline
\textit{server} &  \text{\scriptsize CFE\_PLATFORM\_TIME\_CFG\_SERVER}\\
\hline
\textit{client} & \text{\scriptsize CFE\_PLATFORM\_TIME\_CFG\_CLIENT} \\
\hline
\textit{big\_endian} & \text{\scriptsize CFE\_PLATFORM\_TIME\_CFG\_BIGENDIAN} \\
\hline
\textit{virtual\_met} & \text{\scriptsize CFE\_PLATFORM\_TIME\_CFG\_VIRTUAL} \\
\hline
\textit{source} & \text{\scriptsize CFE\_PLATFORM\_TIME\_CFG\_SOURCE} \\
\hline
\textit{source\_met} & \text{\scriptsize CFE\_PLATFORM\_TIME\_CFG\_SRC\_MET} \\
\hline
\textit{source\_gps} & \text{\scriptsize CFE\_PLATFORM\_TIME\_CFG\_SRC\_GPS} \\
\hline
\textit{source\_time} & \text{\scriptsize CFE\_PLATFORM\_TIME\_CFG\_SRC\_TIME} \\
\hline
\textit{signal} & \text{\scriptsize CFE\_PLATFORM\_TIME\_CFG\_SIGNAL} \\
\hline
\textit{tai} & \text{\scriptsize CFE\_MISSION\_TIME\_CFG\_DEFAULT\_TAI} \\
\hline
\textit{utc} & \text{\scriptsize CFE\_MISSION\_TIME\_CFG\_DEFAULT\_UTC} \\
\hline
\textit{fake\_tone} & \text{\scriptsize CFE\_MISSION\_TIME\_CFG\_FAKE\_TONE} \\
\hline
\textit{at\_tone\_was} & \text{\scriptsize CFE\_MISSION\_TIME\_AT\_TONE\_WAS} \\
\hline
\textit{at\_tone\_will\_be} & \text{\scriptsize CFE\_MISSION\_TIME\_AT\_TONE\_WILL\_BE}\\
\hline
\end{tabular}
\label{var2code}
\end{center}
\vspace*{-4mm}
\end{table}

\begin{table} [tb]
\caption{TIME Build Test and Unit Test Results.}
\vspace*{-3mm}
\begin{center}
\begin{tabular}{|c|c|c|c|}
\hline
\textbf{Build Test}&\multicolumn{2}{|c|}{\textbf{Results}} & \textbf{Total}\\
\cline{2-3} 
\textbf{Component} & \textbf{Pass}& \textbf{Fail} & \textbf{Configurations} \\
\hline
cFS Time Service & 48 & \textcolor{red}{64} & 112  \\
\hline
cFS Time Unit Test & 36 & \textcolor{red}{76} & 112  \\
\hline
\end{tabular}
\label{TestResult}
\end{center}
\vspace*{-4mm}
\end{table}

Table \ref{TestResult} shows that 76 out of the 112 valid configurations failed the build tests for the unit tests, where 64 of the failures were the same as the system's failed build tests. After investigation, all the failures of both system and unit test builds  were found to be caused by three variability bugs. TIME attempts to call {\em unimplemented abstract functions} (\textit{OS\_SelectTone, OS\_SetLocalMET or OS\_GetLocalMET}) when either a selection for the tone signal {\small \textit{(cFE\_PLATFORM\_TIME\_CFG\_SIGNAL=true)}} or for the local hardware MET {\small \textit{(cFE\_PLATFORM\_TIME\_CFG\_VIRTUAL=false)}} is configured. Additionally, there was a unit test coding bug (TIME\textminus6 {\small \textit{(cFE\_PLATFORM\_TIME\_CFG\_CLIENT=true)}}) that was found to occur when the Time Client is configured. 

These build failures indicate that the code does not satisfy the following variability requirements (refer to Table \ref{processedReq} for more information):
\begin{itemize}
\item TIME\textminus9 {\small \textit{(cFE\_PLATFORM\_TIME\_CFG\_VIRTUAL=false)}}
\item TIME\textminus15 {\small \textit{(cFE\_PLATFORM\_TIME\_CFG\_SIGNAL=true)}}
\item TIME\textminus6 {\small \textit{(cFE\_PLATFORM\_TIME\_CFG\_CLIENT=true)}} in unit test code only 
\end{itemize}

\subsection{Research Questions} \label{researchQ}
We now discuss what our findings from application of the VarCORE on the cFE indicate in terms of the research questions.  

{\bf RQ1: Are there missing variability-related requirements and/or constraints?}

Several variability requirements and constraints are missing from both the cFE Software Requirement Specification (cFE requirements.docx) and the cFE Funtional Requirements document (cFE FunctionalRequirements.csv). During variability requirements and constraints gathering, we found that together they only describe 6 of the  variability requirements for the TIME module compared to 17 (10 variability requirements and 7 constraints) that we extracted from the cFE User's Guide. There is currently no centralized requirement document from which all behavioral requirements and constraints could be identified and understood. 

{\bf RQ2: Are there inconsistencies among the specified
variability requirements and/or constraints?}

Review of the RTW identified a variability requirement inconsistency during the review process. TIME\textminus7 (see Table \ref{processedReq}) cannot be satisfied as there is no definition for processor count in TIME nor imported to TIME. We also found through the feature model analysis that the constraint  TIME\textminus11 conflicted with TIME\textminus10 and TIME\textminus16. These issues require requirements updates to address the inconsistencies.

{\bf RQ3: Are there variability requirements and/or constraints not implemented in the code?}

We investigated this question by testing all 112 valid configurations (product variants) using cFS's existing build system. The results show that TIME\textminus9 and TIME\textminus15 are not implemented in the code. Moreover, there is a variability bug found in unit test for supporting TIME\textminus6. All three of these variability bugs are known open issues tracked by cFS's Bug ID \#109 ( https://github.com/nasa/cFE/issues/109). 

We analyzed the history of resolved bugs for TIME and found a variability bug \#2072 (https://github.com/nasa/cFE/issues/2072) which occurred only for a specific TIME configuration, one of the valid 112 product variants generated by VarCORE. We reproduced the bug (temporarily undoing the fix) when running the build test using the specific product variant as TIME configuration and confirmed that the variability requirement was correct but not implemented correctly in the code. This resolved bug could have been detected earlier if VarCORE had been integrated into the cFS build system.

During code review for answering RQ3, we also found that the specified constraint described in the RTW for TIME\textminus4 is inconsistent with the code implementation. TIME\textminus4 specifies that the issuance of a fake tone shall be enabled only for hardware configurations that do not provide a tone signal. Since both local hardware MET and external source MET can provide a tone signal, the fake tone should be disabled in these two configurations. However, the  implementation of this variability allows a fake tone when the local hardware MET is enabled. The implemented constraint is inconsistent with the specified constraint. 

\section{Discussion}
\subsection{Lessons Learned}\label{lessonlearned}

We report four lessons learned from our experience with the flight software that may be useful on other industrial projects. 
\begin{itemize} 
\item
Information about variability-related requirements and constraints may be dispersed among documents, not all of which are labeled as requirements.  This makes it more time-consuming to identify and understand optional requirements and constraints on the design space.  It also makes it easier to miss needed requirements or to inadvertently violate constraints by introducing conflicting requirements.  

\item Variability-related requirements and constraints are easily overlooked and need special attention in mission-critical software. The flight software framework uses many best practices for variability management, as described in Sect. \ref{context}, and is well maintained by experienced developers. For a new project with developers initially unfamiliar with the flight software framework, there remain obstacles to navigating the maze of information \cite{CashmanCRC18} to determine their project requirements.  Our findings (Section \ref{App_Findings}) suggest that special attention to variability-related issues  can simplify future developers' tasks.    
\item Lightweight variability modeling, as in VarCORE, appears to be effective in finding errors. We identify two reasons for this.  (1) It focuses special attention on the variability. (2) It centralizes the dispersed requirements in a structured format (the RTW) that enables automated creation and analysis of a standard feature model. 

\item A graphical view of the variability requirements and constraints makes it easier to see and understand inconsistencies. VarCORE uses a well-known tool (FeatureIDE) to display the feature model in a tree-based, graphical display annotated with any constraint violations, and to generate all valid configurations in support of build and unit testing.  

\end{itemize}

An interview-based study 2021  by Schmid et al. noted that ``industry still struggles to deal with a high number of variants of their systems systematically. The underlying issue, i.e., that knowledge about variability is often only tacit, available from the heads of the developers only, has not disappeared'' \cite{SchmidR0BGGW21}.  Similarly, Kruger et al. reported in a 2019 study that the availability of lightweight traceability of features to source code immediately benefited both developers and maintainers \cite{KrugerCBLS19}. We sought in our work to improve the state of practice by making the knowledge about constraints on valid feature combinations less tacit and more obvious to the developers of a new flight software system using the cFS framework.  

\subsection{Threats to Validity}\label{threats}
An internal threat to validity is that, while verification that our feature model is sound uses the static analysis performed by the FeatureIDE tool to confirm that it produces only valid configurations, the variability requirements or constraints themselves may be incorrect. VarCORE can assist to some extent in surfacing such errors since it checks that all the configurations identified as valid by FeatureIDE can be compiled and built for unit testing.  Another threat to internal validity is that   several steps in VarCORE are manual (requirement/constraint extraction, propositional logic conversion) and require domain expertise, with results dependent on the accuracy of input provided by developers. We can automate more of the flow to lessen this dependency; however, the current worksheet-to-model approach has the benefit of being familiar with a low bar to adoption.  Our selection of one project, the NASA flight software, for application may affect the external validity.  However, we chose it due to our  experience with it, and it has a sizeable community that has used it on a variety of aerospace projects.  

\subsection{Related Work} \label{relatedwork}
Variability models are well-studied and take multiple forms including feature models \cite{kang1990feature}, tabular configurability models \cite{CashmanCRC18}, and formal models \cite{cChechik18}, created either from requirements or reverse engineered from code. 

Automating requirements and feature extraction using NLP remains a challenging task and requires domain expertise to fine tune the tools and verify the results. 
At RE'22 Rajbhoj et al. described successful extraction of traceability information from requirements specifications that are highly structured, which was not the case here \cite{RajbhojNKSP22}. 

Our approach is perhaps most similar to that described by Acher et al., who also proposed a tool-supported process to create a feature model from tabular specifications\cite{acher2012extracting}.  They assumed input was already in tabular form before normalizing these requirements into a more structured format for interpretation and synthesizing the feature model based on high-level defined directives. In contrast, we gather the variability requirements into a central Requirements Traceability Worksheet (RTW) for metadata augmentation to facilitate traceability between the requirements and the feature model. We also leverage a popular open-source feature modeling tool for constraint analysis, which flight software developers are more likely to accept than an unfamiliar tool.  

More broadly, Gazzillo and Cohen recently urged that configurability be promoted to a first class element, noting that ``configurable software makes up most of the software in use today." They cite a need for common ground to ``bring together researchers and practitioners who are typically siloed" \cite{GazzilloC22}. We hope that the open-source NASA flight software framework may provide one such opportunity for joint innovation.  

\subsection{Future Work} \label{futurework}
A long-range goal is to include VarCORE in the cFS tool suite.  Toward this goal we plan to automate additional portions of VarCORE, perhaps using existing NLP techniques \cite{abualhaija2019machine,nistala2022towards,zhang2022automatic}, and to enable variability code auto-analysis \cite{patterson2022sugarc,schubert2022static} to evaluate the consistency between requirements specification and implementation. 

\section{Conclusion}
\vspace*{-1mm}
This paper has described our experience with the variability-related requirements and requirement constraints for a configurable NASA flight software framework.   A challenge for new projects using this framework is that the specification of variability requirements and constraints are dispersed across multiple documents. This can contribute to requirements inconsistencies, omissions, and conflicts, as well as to inadvertent violations of constraints on combinations of options.   We described VarCORE, a new structured technique to create a variability model for the  configurable flight software using feature-modeling and open-source tools. We showed in our application of it to a critical piece of the flight software how VarCORE centralizes, specifies, and analyzes the variability requirements and constraints that are currently dispersed.  We reported its effectiveness in finding some missing and inconsistent variability requirements and constraints. Finally, we discussed the broader lessons learned from our experience that may be useful to requirements teams on other large configurable industrial projects.

\section*{Acknowledgment}
\vspace*{-1mm}
We thank the reviewers for their helpful feedback. 

\bibliographystyle{IEEEtran}
\bibliography{Reference/References.bib}

\end{document}